\documentclass[useAMS,usenatbib]{mn2e}
\usepackage{graphicx}
\usepackage{times}

\newcommand{\Ms}{$M_{\star}$}

\usepackage{savesym}
\usepackage{amsmath}
\savesymbol{iint}
\usepackage{txfonts}
\restoresymbol{TXF}{iint}

\bibpunct{(}{)}{;}{a}{}{,}

\title[GAMA: metals and HI content in galaxies]{Galaxy And Mass Assembly (GAMA): The connection between metals, specific-SFR, and HI gas in galaxies: the Z-SSFR relation}
\author[M. A. Lara-L\'opez]{M. A. Lara-L\'opez$^{1}$\thanks{E-mail:
mlopez@aao.gov.au},  A. M. Hopkins$^{1}$, A. R. L\'opez-S\'anchez$^{1,2}$, S. Brough$^{1}$, M. Colless$^{1}$
\newauthor J. Bland-Hawthorn$^{3}$, S. Driver$^{4,5}$, C. Foster$^{6}$, J. Liske$^{7}$, J. Loveday$^{8}$,  A. S. G. Robotham$^{4,5}$
\newauthor  R. G. Sharp$^{9}$, O. Steele$^{10}$, E. N. Taylor$^{3,11}$\\
$^{1}$Australian Astronomical Observatory, PO Box 915, North Ryde, NSW 1670, Australia\\
$^{2}$Department of Physics and Astronomy, Macquarie University, NSW 2109, Australia.\\
$^{3}$Sydney Institute for Astronomy (SIfA), School of Physics, University of Sydney, NSW 2006, Australia\\
$^{4}$International Centre for Radio Astronomy Research, The University of Western Australia, 35 Stirling Highway, Crawley, WA 6009, Australia\\
$^{5}$School of Physics \& Astronomy, University of St Andrews, North Haugh, St Andrews, KY16 9SS, UK\\
$^{6}$European Southern Observatory, Alonso de Cordova 3107, Vitacura, Santiago, Chile\\
$^{7}$European Southern Observatory, Karls-Schwarzschild-Str. 2, 85748 Garching, Germany\\ 
$^{8}$Astronomy Centre, University of Sussex, Falmer, Brighton BN1 9QH\\
$^{9}$Research School of Astronomy \& Astrophysics, Australian National University, Cotter Road, Weston Creek, ACT 2611, Australia\\
$^{10}$Institute of Cosmology and Gravitation, University of Portsmouth, Dennis Sciama Building, Burnaby Road, Portsmouth PO1 3FX, UK\\
$^{11}$School of Physics, The University of Melbourne, Parkville, VIC 3010, Australia\\}

\begin{document}

\date{Accepted  . Received   ; in original form    }

\pagerange{\pageref{firstpage}--\pageref{lastpage}} \pubyear{2013}

\maketitle

\label{firstpage}

\begin{abstract}

We study the interplay between gas phase metallicity (Z), specific star formation rate (SSFR) and neutral hydrogen gas (HI)  for galaxies of different stellar masses. Our study uses spectroscopic data from GAMA and SDSS star forming galaxies, as well as HI--detection from the ALFALFA and GASS public catalogues. We present a model based on the Z-SSFR relation that shows that at a given stellar mass, depending on the amount of gas, galaxies will follow opposite behaviours. Low--mass galaxies with a large amount of gas will show high SSFR and low metallicities, while low--mass galaxies with small amounts of gas will show lower SSFR and high metallicities. In contrast, massive galaxies with a large amount of gas will show moderate SSFR and high metallicities, while massive galaxies with small amounts of gas will show low SSFR and low metallicities. Using ALFALFA and GASS counterparts, we find that the amount of gas is related to those drastic differences in Z and SSFR for galaxies of a similar stellar mass.
\end{abstract}


\section{Introduction}

The formation of galaxies is intimately dependent on the conversion of stars into gas, the production of heavy elements, recycling of this material into the interstellar medium, and repetitions of this cycle. A detailed understanding of the interplay between each of gas mass, star formation rate, and metallicity is clearly important to understand the galaxy evolution process. Scaling relations between SFR and stellar or gas mass, and between mass and metallicity, have been explored for many years. More recently, the connection between  gas fraction and metallicity have also begun to be explored \citet[][]{Hughes12}. Identifying the causal relationships between these empirical scaling relations is, however, more challenging.

% Scaling relations between fundamental properties of galaxies provide crucial information about the mechanisms that drive their evolution. 

From the fundamental properties of galaxies, we can identify extensive (scale-dependent) and intensive (scale-invariant) properties. Intensive properties are essential or inherent to a system and are independent of the amount of baryonic mass, for example metallicity, temperature, density, and SSFR. Extensive properties on the other hand, depend on the amount of baryonic mass of the system, such as the star formation rate (SFR) and stellar mass (\Ms). By way of illustration of this point, if we were to split a galaxy in half, the metallicity of the halves are unchanged from the original galaxy. The SFR and stellar mass of the halves, however, are half that of the original, corresponding to the amount of baryonic mass removed.

% It is important to understand this difference because the fact that 2 variables correlate does not mean that they depend one to the other. 
% A very well known relation for star forming (SF) galaxies that combines extensive and intensive properties is the \Ms$-Z$ relation \citep[e.g.][]{Tremonti04,Lara09a, Lara09b}. 

Although the metallicity is scale invariant, or intensive, it is related to the stellar mass through the well known \Ms$-Z$ relation \citep[e.g.][]{Tremonti04,Lara09a, Lara09b}. In this relationship, the metallicity of high mass systems is higher than that of low mass systems. This does not mean that the metallicity and \Ms\ are dependent variables. A correlation or trend between an intensive and an extensive property cannot be directly causal. For example, simply combining a large number of low-metalliticy dwarf galaxies will increase the total mass, but will not directly increase the metallicity of the total system. Any such relation, consequently, must imply a common underlying cause, in this example likely to be the star formation history of the galaxy.

% Similarly, there is a relation between the SSFR and the \Ms\ of star forming (SF) galaxies \citep[e.g.][Bauer et al. 2012, submitted]{Noeske07}, with low--mass galaxies showing higher SSFRs than high--mass galaxies. 

On the other hand, relationships such as Z-SFR \& Z-SSFR are shown to have weaker correlations \citep[e.g.][]{Lara10a,LopezSanchez10, Lara13a}. Specifically, the Z-SSFR relation is formed by intensive properties of galaxies, and has not been widely studied since it shows  high scatter. Nevertheless, metallicity is important for the ability of gas clouds to form stars, since it enables most of the gas cooling that will facilitate the star formation \citep[][]{Labou13}.

% Among the first studies about the Z-SSFR are \citet{Lara10a}, who although showing a high scatter in this relationship, they also show the possible existence of two branches.

Crucial information regarding galaxy properties is given by their neutral gas content. Large HI surveys such as the Arecibo Legacy Fast Arecibo L-band Feed Array \citep[ALFALFA, ][]{Haynes11}, and the Galex Arecibo SDSS Survey \citep[GASS, ][]{Catinella10}, have provided new scaling relationships and dependencies for the cool gas in galaxies.
% \citet{Catinella12} study the HI properties of selected massive ( $>$10$^{10}$\Ms) SDSS galaxies. They derive a \textit{gas fraction plane} \citep[see also][]{Catinella10}, which is a relation between gas mass fraction and a linear combination of NUV-r colour and stellar mass surface density.

The link between the gas metallicity and HI  content in galaxies was studied in \citet{Zhang09}. Specifically, they  study the HI  dependence on the \Ms$-Z$ relation using SDSS galaxies. They estimate empirically the HI  content for 10$^5$ emission--line galaxies in the SDSS-DR4, and find that gas--poor galaxies are more metal rich at fixed stellar mass.  \citet[][]{Hughes12} confirms this result, from an investigation of  the role of cold gas and environment on the \Ms$-Z$ relation for 260 nearby late--type galaxies. They find that, at fixed stellar mass, galaxies with lower gas fractions typically also possess higher metallicities. In general, they observe that gas--poor galaxies are typically more metal rich, and demonstrate that the removal of gas from the outskirts of spirals increases the observed average metallicity by $\sim$ 0.1 dex.

Athough some scaling relations with HI content have been studied, the interplay between  metallicity and SSFR has not yet been analyzed as a function of the gas content or stellar mass. The Z-SSFR relation is formed by intensive properties, and is a key relationship to study the different properties of high--  and low--mass galaxies.

The gas mass is a direct measure of the available fuel in galaxies to form stars. The relation of this gas with metallicity, stellar mass, and SSFR can tell us how fast a galaxy assembled stars in the past, the amount of stars it formed, and how actively it is forming stars at present. The main goal of this paper is to analyse the interplay between all these properties to produce a general picture of the gas recycling process.

% does not depend on stellar mass.

This paper is organized as follows. In $\S\,\ref{SampleSelection}$ we detail the data used for this study, and in   $\S\,\ref{ZSSFR}$ we  analyse the Z-SSFR relation and present a cartoon model based on it. In  $\S\,\ref{HIScalRel}$  we study scaling relations using HI data. Finally, in $\S\,\ref{Conclusion}$  we present our discussion and conclusions. Throughout we assume $H_0=70\,$km\,s$^{-1}$\,Mpc$^{-1}$, $\Omega_M=0.3$, $\Omega_{\Lambda}=0.7$.

%__________________________________________________________________

\section[]{Sample selection}\label{SampleSelection}

\subsection[]{Optical data}\label{OpticalData}

We consider data for emission-line galaxies from two large surveys, the Galaxy and Mass Assembly \citep[GAMA phase-I survey, ][]{Driver11}, and the  Sloan Digital Sky Survey--Data Release 7 \citep[SDSS--DR7,][]{Abaza09}.

Data from the SDSS were taken with the 2.5 m telescope located at Apache Point Observatory \citep{Gunn06}. We use the emission-line analysis of SDSS-DR7 galaxy spectra performed by the MPA-JHU  database\footnote{http://www.mpa-garching.mpg.de/SDSS/}. From the full dataset, we only consider objects classified as galaxies in the $``$main galaxy sample$"$ \citep{Strauss02} with apparent Petrosian $r$--magnitudes in the range $14.5 < m_{r} < 17.77$ and $z<0.33$. We use gas metallicities measured as described in \citet{Tremonti04}, SFR estimates described in \citet{Brinchmann04}, and total stellar masses estimated as in \citet{Kauf03a}.

% basically, they use estimated metallicities statistically using Bayesian techniques based on simultaneous fits of all the most prominent emission lines ([{O\,\textsc{ii}}], {H$\beta$}, [{O\,\textsc{iii}}], {H$\alpha$}, [{N\,\textsc{ii}}], [{S\,\textsc{ii}}]), using a model designed for the interpretation of integrated galaxy spectra \citep{Charlot01}. 
% 
% SFRs estimates are described in \citet{Brinchmann04}, who are based also on Bayesian methods. \citet{Brinchmann04} estimated SFRs modeling the emission lines in the galaxies following the  \citep{Charlot02} prescription, achieving a robust dust correction. 

% Total stellar masses were estimated as in \citet{Kauf03a}, which relies on spectral indicators of the stellar age, and the fraction of stars formed in recent bursts.  These authors used the $z-$band magnitude to characterize the galaxy luminosity and constrained the star formation history using both the 4000\AA\ break, $D_n$(4000), and the stellar HI Balmer absorption, H$\delta_A$. The location of a galaxy in the $D_n$(4000)--H$\delta_A$ plane is insensitive to reddening and it depends weakly on metallicity. 

The GAMA phase-I is a spectroscopic survey  with the 3.9m Anglo-Australian Telescope (AAT) using the 2dF fibre feed and AAOmega multi-object spectrograph. For full details of the survey selection and properties see \citet{Driver11}.  SFRs measurements are based on the {H$\alpha$} emission line as described in  \citet{Mad11}. Metallicities were estimated  using the  empirical calibration  provided by  \citet{Pettini04}  between the oxygen abundance and the O3N2 ( ([${\rm O\,\textsc{iii}}] \;  \lambda 5007/{\rm H}\beta) / ([{\rm N\,\textsc{ii}}]  \; \lambda 6583/{\rm H}\alpha$) )  index. Both SFRs and metallicities were recalibrated to the Bayesian system described above using the calibrations of \citet{Lara13b}. Finally, stellar masses were measured as described in \citet{Taylor11}.
% , who are consistent with \citet{Kauff03a}.

For both surveys, we selected only SF galaxies using the standard spectroscopic diagnostic \citep{Baldwin81}, and using the discrimination of \citet{Kauf03b}. For reliable metallicity and SFR estimates, we selected galaxies with a signal-to-noise ratio (SNR) of 3 in  {H$\alpha$}, {H$\beta$}, and  [{N\,\textsc{ii}}]. Galaxies were selected in volume--limited samples in redshift and Petrosian r-band absolute magnitude as described in \citet{Lara13b}. Our final SF optical sample is 35212 galaxies for GAMA and 156910 for SDSS. These samples are used to match with ALFALFA and GASS data as described in the next section.

\subsection[]{HI data}\label{HIdata}

To test the correlations between optically derived properties and the HI content of galaxies, we used the ALFALFA \citep[][]{Haynes11} and the GASS \citep[][]{Catinella10} Surveys.

ALFALFA is a blind survey of 21 cm HI emission  over $\sim$ 2800 deg$^2$ of sky. The public catalogue\footnote{http://egg.astro.cornell.edu/alfalfa/data/index.php} provides $\sim$ 15,855 HI detections, of which $\sim$ 15,041 are associated with extragalactic objects. Since ALFALFA is a blind survey, we select only galaxies with Code=1, which refers to robust, reliably detected sources. We also remove sources with heliocentric velocities V$_{helio}$$<$100.0, which are unlikely to be galaxies. From this subsample, we cross-match RA, DEC, and redshift for our optical SF--sample described in $\S\,$\ref{OpticalData}, obtaining a final sample of 4443 SF galaxies with ALFALFA counterparts. These come entirely from the SDSS, since the GAMA survey regions are not well covered by the publicly available ALFALFA data. Together with the fainter magnitude range of the GAMA targets, this biases against them being identified with ALFALFA detections.

In order to increase our HI coverage of massive galaxies, we match our SDSS SF--sample described in  $\S\,$\ref{OpticalData} with the GASS public catalogue. The GASS survey aims to observe a sample of $\sim$ 1000 galaxies selected from the SDSS spectroscopic and GALEX imaging surveys. GASS galaxies are selected to have stellar masses greater than 10$^{10}$ \Ms\ and redshifts 0.025$<z<$0.05. The public catalogue includes 232 galaxies, from which we find 48 counterparts with our optical  SF--sample described in $\S\,$ \ref{OpticalData}. The rest of the GASS galaxies correspond to composite and AGN galaxies. The stellar mass range of our final sample is $10^7<$\Ms$<10^{11}$ M$_{\sun}$, with a median of $\sim$  $10^{9.3}$M$_{\sun}$.%, as shown in Fig. \ref{HistMass}.

% Although we are including GASS galaxies to increase our high mass galaxy sample, the histogram of Fig. \ref{HistMass} shows that in our final sample we do not have galaxies with \Ms $>$ 10$^{11}$. 
% The median redshift of our final sample is 0.04.

% Unfortunatelly since GAMA regions are not well-covered by ALFALFA we did not find counterparts with GAMA. Also, since GAMA probes lower-mass and higher redshift galaxies than SDSS, there is a little likelihood of finding ALFALFA counterparts in GAMA.
% 
% \begin{figure}
% \begin{center}
% \includegraphics[scale=0.25]{HistStellarMassHIgasSample.ps}
% \caption{Stellar mass histogram of our final sample with HI  measurements including both ALFALFA and GASS detections.}
% \label{HistMass}
% \end{center}
% \end{figure}

\section[]{The Z-SSFR relation}\label{ZSSFR}

The correlation between the amount of metals and the SSFR has not been explored in detail. Since both Z and SSFR are intensive properties, this relationship provides a direct way to analyse the different properties of low-- and high--mass galaxies. 
%  \citet{Lara10a} used SDSS data to observe signs of two possible tails in this relationship separated at log(SSFR) $\sim -10$ yr$^{-1}$.
% Recently, \citet{Yates12} studied the Z-SSFR relation through the median Z in bins of SSFR and \Ms. However, different binning directions provide different information, and correlations with high scatter can be very sensitive to this. See further discussion in  \citet{Lara13a} and  \citet{Lara13b}.

Here we explore the median SSFR in bins of Z for different \Ms. In Fig. \ref{DependenciaBin1p1} we show the Z-SSFR relation for SDSS and GAMA data. Each of the panels correspond to different redshifts of volume--limited samples as described in  \citet{Lara13b}. The coloured circles correspond to the median SSFR in bins of Z for different \Ms. Opposing trends are  clearly observed  in Figs. \ref{DependenciaBin1p1}a and \ref{DependenciaBin1p1}b   between low-- and high--mass galaxies with an inflection point at log(SSFR) $\sim-$9.9  yr$^{-1}$. The SSFR for low--mass galaxies shows an anticorrelation with Z while the SSFR for high--mass galaxies correlates with Z.

\subsection[]{A  Model}\label{CModel}

\begin{figure*}
\includegraphics[scale=0.87]{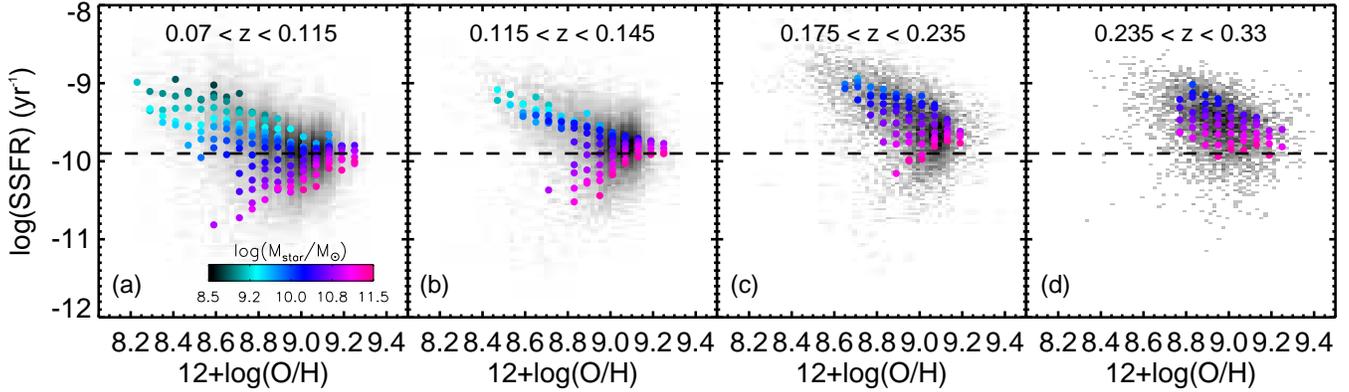}
\caption{Z-SSFR relation for SDSS and GAMA data.The colour coded circles correspond to the median  SSFR in bins of Z for different bins of \Ms. The grey points correspond to the density of SDSS and GAMA samples.}
\label{DependenciaBin1p1}
\end{figure*}

\begin{figure*}
\includegraphics[scale=0.60]{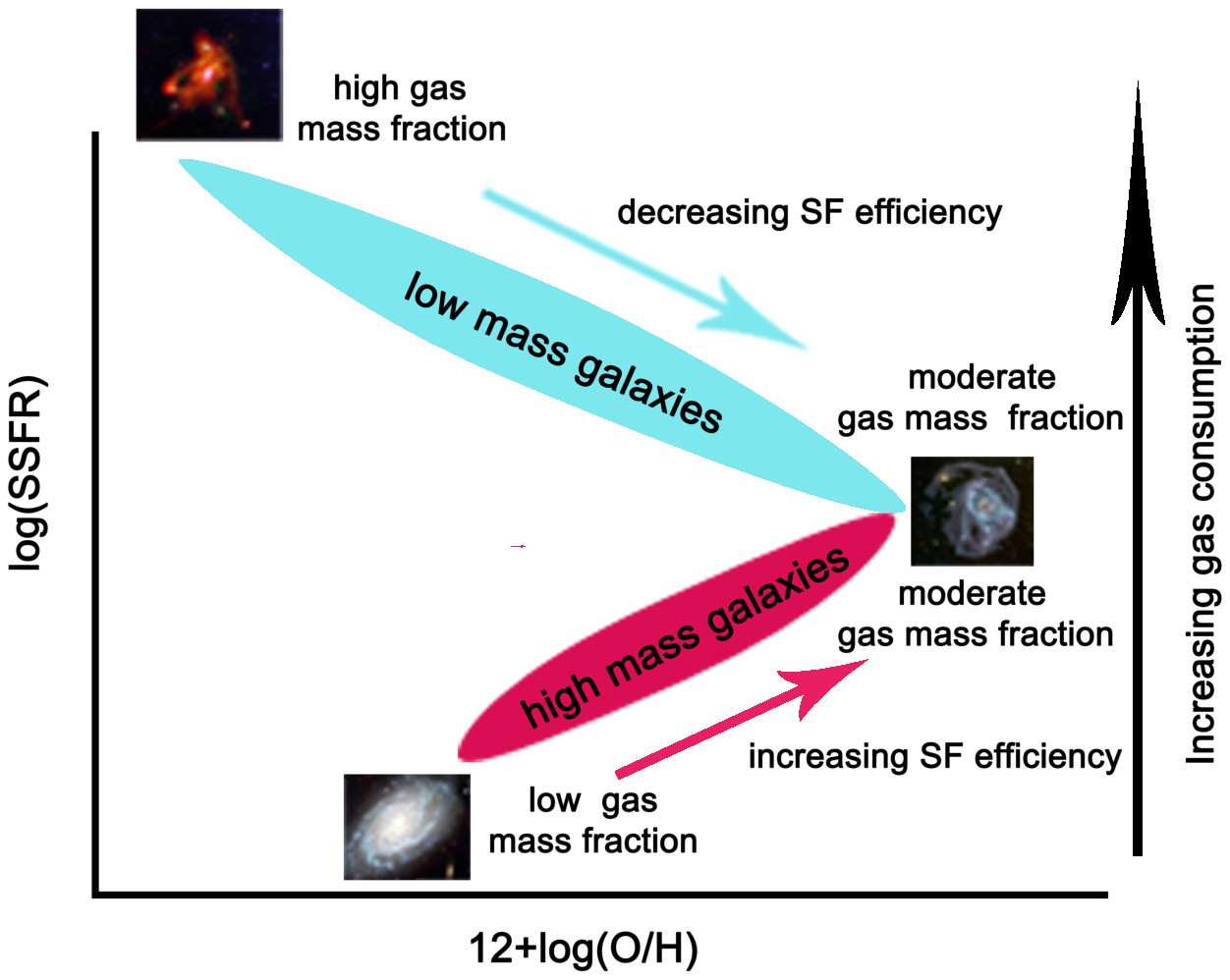}
\includegraphics[scale=0.45]{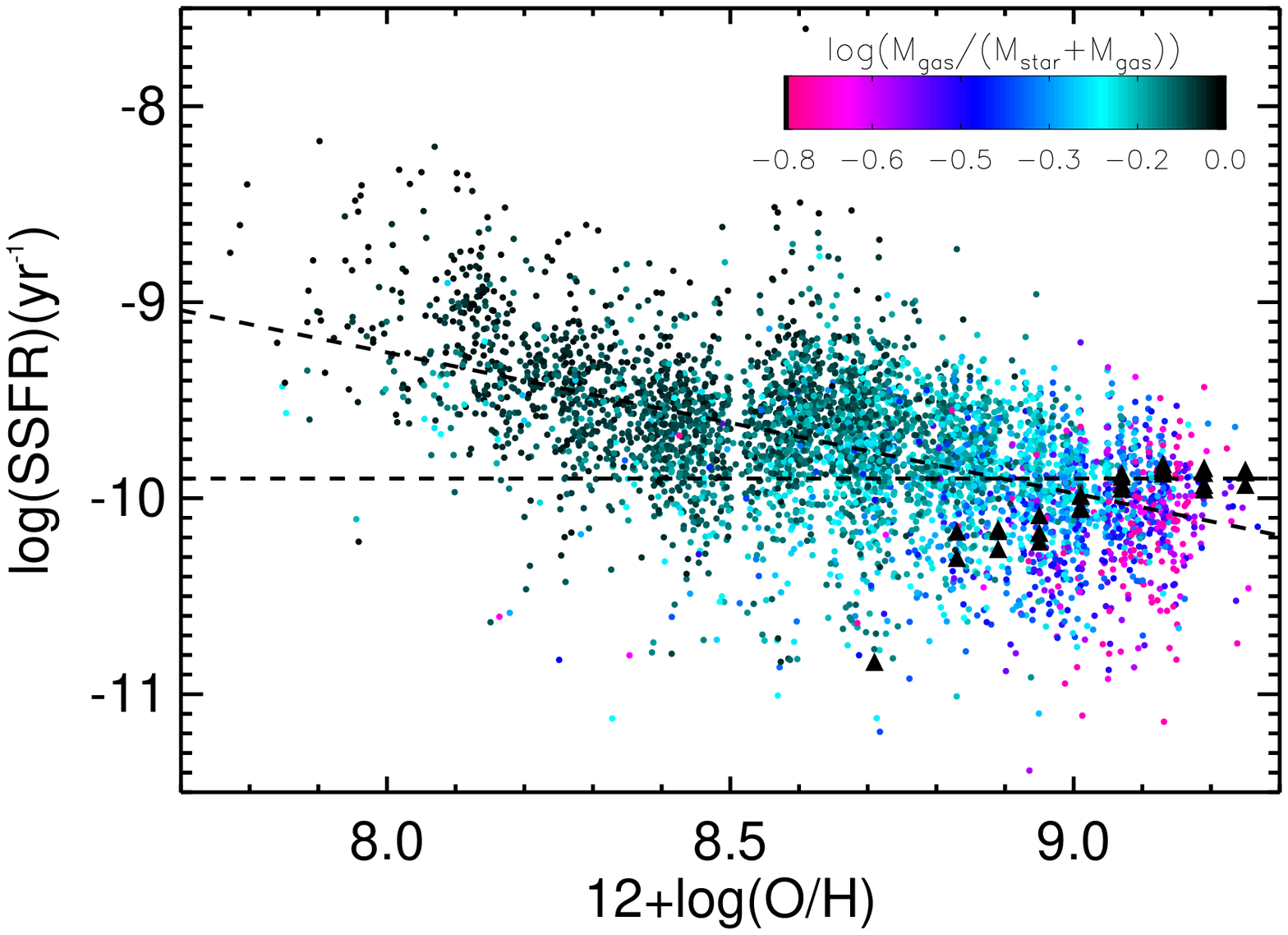}
\caption{Left: cartoon model of the $Z-$SSFR relationship  based on the extreme stellar mass cases of Fig. \ref{DependenciaBin1p1}. The model is based on local redshift galaxies, and represents a snapshot of the behaviour of galaxies today. The blue and red ellipses represent low-- and high--mass galaxies, respectively. Right: $Z-$SSFR relation for galaxies with ALFALFA and GASS counterparts. Galaxies are colour coded from low (magenta) to high (green)  gas mass fraction. Black triangles show the median data for SDSS galaxies with no HI detection in the ALFALFA and GASS fields for  log(\Ms) $>$ 11.0 dex.}
\label{CartoonModel}
\end{figure*}

\begin{figure*}
\includegraphics[scale=0.33]{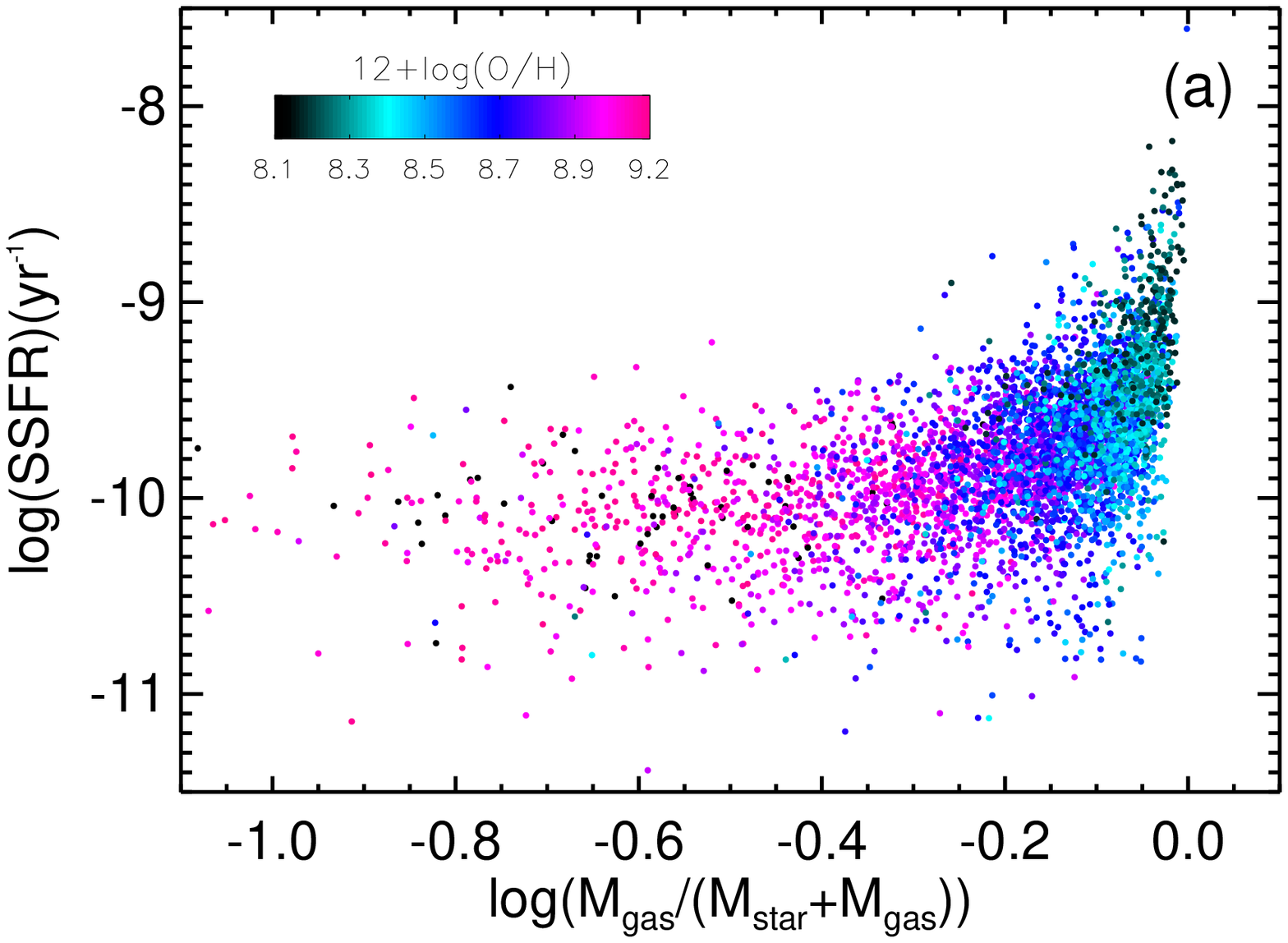}
\includegraphics[scale=0.35]{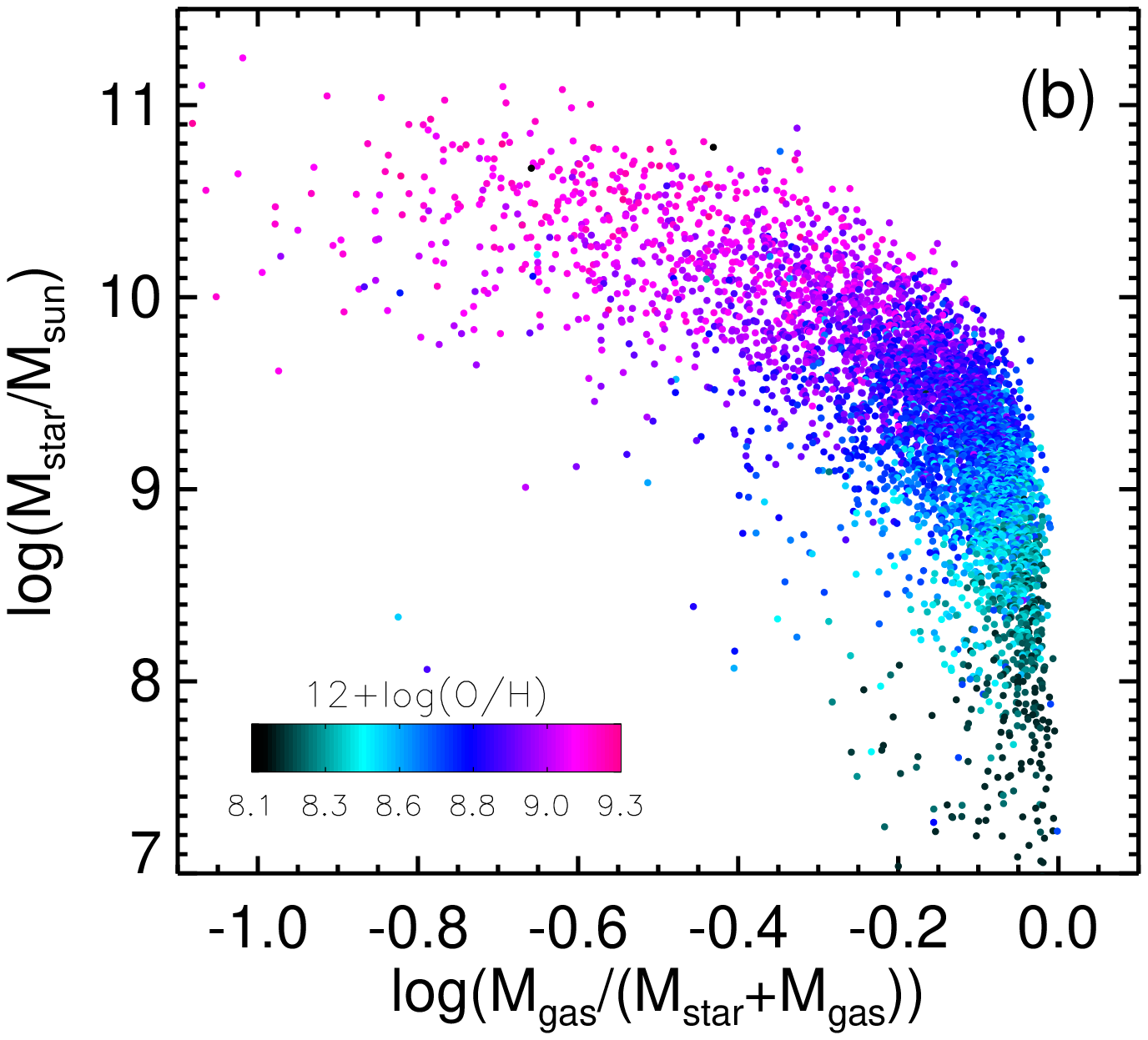}
\includegraphics[scale=0.35]{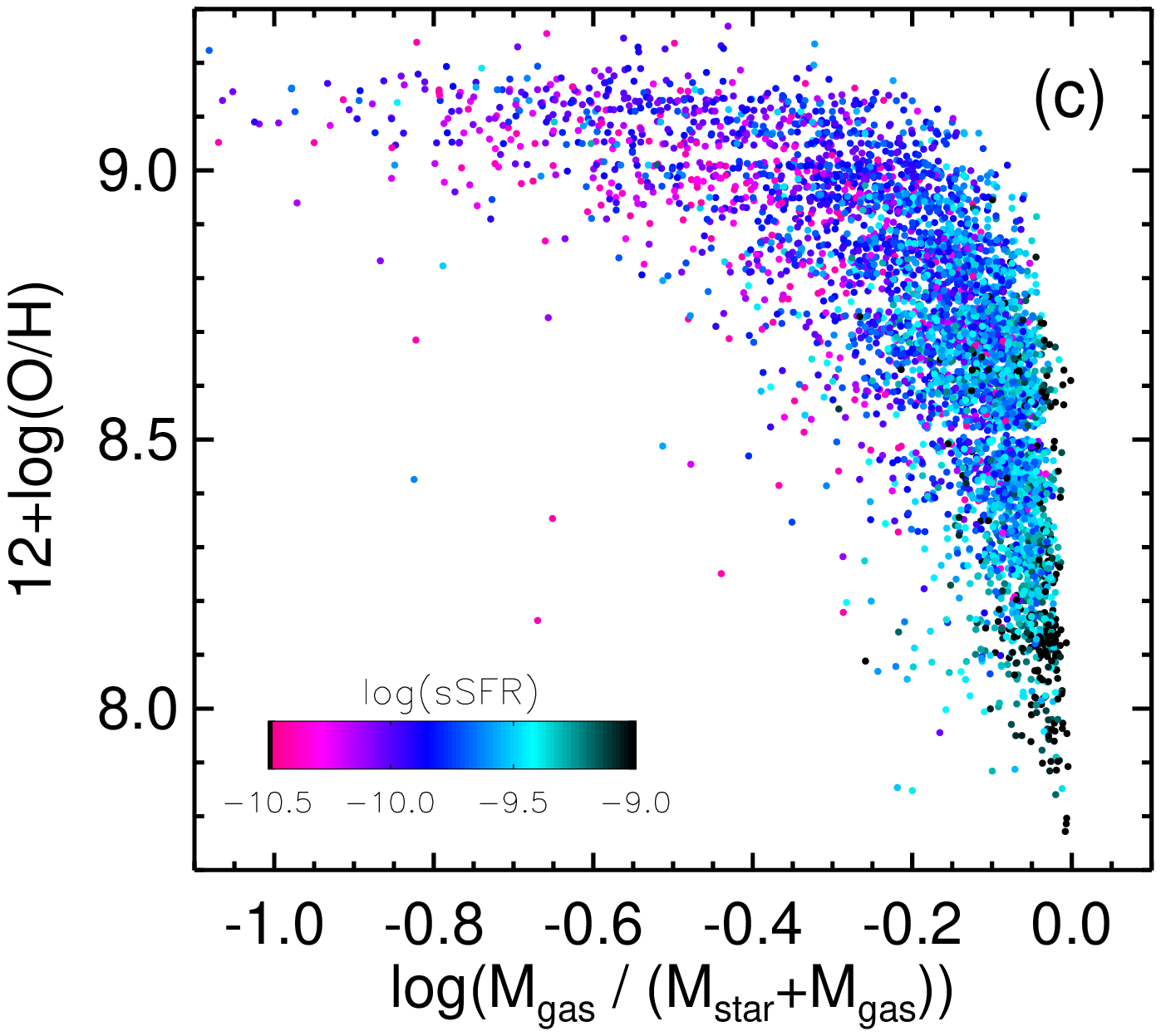}
\caption{Scaling relations for SDSS galaxies with ALFALFA and GASS counterparts. (a) The gas mass fraction and SSFR relation colour coded from low (green) to high (magenta) metallicity. (b) The gas mass fraction and \Ms\ relation colour coded as in Figure (a). (c) The gas mass fraction and Z relation for the same sample colour coded from low (magenta) to high (green) SSFR.}
\label{HIgas}
\end{figure*}

We suggest that the opposite behaviors for low and high mass galaxies observed in Fig. \ref{DependenciaBin1p1} can be understood assuming that galaxies at a given \Ms\ have different amounts of HI. Based on that figure, we generate the cartoon model shown in Fig. \ref{CartoonModel}. In this model, low--mass galaxies are represented by a blue ellipse that extends from high SSFR and low $Z$, to low SSFR and high $Z$. Massive galaxies, represented by a red ellipse, extend from low SSFR and low $Z$ to high SSFR and high $Z$.

Different physical explanations can be given to justify this basic supposition of our cartoon model. The presence of galactic inflows and outflows and how they affect the metal enrichment in galaxies is still a matter of debate. Galactic outflows have been shown to be important to reproduce the \Ms$-Z$ relation of galaxies using analytical \citep[e.g.,][]{Erb08} and hydrodynamic models \citep[e.g.,][]{Fin08}. On the other hand, \citet{Calura09} reproduced the \Ms$-Z$ relation mainly by means of an increasing efficiency of star formation, without any need to invoke galactic outflows.

A general picture of downsizing suggests that low--mass galaxies  process their gas slower than massive galaxies, therefore low--mass galaxies are actively forming stars, and  present high SSFRs today compared to massive galaxies. Hereafter, we will focus on downsizing to provide a possible explanation for the trends between the intensive properties of SSFR and Z as a function of the extensive property of mass.

A combination of downsizing and the amount of HI  can explain the opposing trends  for low and high mass galaxies observed in the Z-SSFR relation. Focusing our attention on the low--mass galaxy branch (blue ellipse of Fig. \ref{CartoonModel}), we can explain the differences in SSFR  through the amount of HI  present in each galaxy. Low--mass galaxies with a high amount of neutral gas will show higher SSFR today, as they have more fuel for star formation, than  galaxies of the same mass  with a lower amount of neutral gas. The same is applicable to high--mass galaxies (red ellipse of Fig. \ref{CartoonModel}), in this branch, galaxies with a high amount of HI  show  higher SSFRs than galaxies at the same mass  with lower amount of HI.

It is noteworthy however, that $Z$ plays the opposite role. While low--mass galaxies with a large amount of HI  show low $Z$, massive galaxies with a large amount of HI  show high $Z$. This can again be explained by downsizing, but not downsizing in stellar mass, rather the amount of HI is driving the rate of metal enrichment in galaxies. In the low--mass branch, galaxies with large  HI  show low $Z$ because they are processing their gas slower and on longer time scales than galaxies with lower  HI  for the same stellar mass. On the other hand, massive galaxies with  large HI show higher metallicities because they processed their gas faster in the past and  have already reached high $Z$, and due to their large HI, they also have a high SSFR too.

% {\bf Comparing with other recent work, \citet{Yates12} find an analogous reverse in the sense of the \Ms$-Z$ relation as a function of stellar mass (their Fig. 1). In this figure, it can be appreciated that high-- and low--mass galaxies also show opposite trends. For massive galaxies, the metallicity is higher for galaxies with high SFR than for galaxies with low SFR, and opposite for low mass galaxies. Our model also explains  this reverse of the high-- and low--mass end in the \Ms$-Z$ relation.}

% According to Fig. \ref{CartoonModel}, for a given stellar mass, galaxies exhibit a wide range of SSFRs. The SSFR for  high--mass  goes from low to high values, and correlates with metallicity (the red arrow in the cartoon model). On the other hand, the low--mass ellipse goes from low to high SSFRs, but anticorrelates with $Z$ (the blue arrow in the cartoon model). This opposite behaviour at the low-- and high--mass ends will result in a reverse of the \Ms$-Z$, \Ms--SFR, and \Ms--SSFR relations, as shown in \citet{Yates12} and \citet{Lara13b}.

\section[]{HI scaling relations}\label{HIScalRel}

To test our model we use the 4491 SDSS optical counterparts with direct HI mass measurements from the ALFALFA and GASS surveys described in $\S\,$\ref{HIdata}. For the relationships shown in this section, we define the gas mass as M$_{\rm{gas}}=1.32\times$M$_{\rm{HI}}$, and the gas mass fraction as  M$_{\rm{gas}}$ / ($\rm{M_{\star}}$ + M$_{\rm{gas}}$).

We generate the Z-SSFR relation and show it as a function of the gas mass fraction in Fig. \ref{CartoonModel} (right). Although with our HI sample it is only possible to see the low--mass branch in this relationship, it is clear that the gas mass fraction of galaxies increases as the SSFR increases, consistent with our proposed cartoon model. Black triangles in Fig. \ref{CartoonModel} (right) correspond to SDSS galaxies in the ALFALFA and GASS fields with no HI detection and \Ms $>$10$^{11}$ M$_{\sun}$. A linear fit to these data is given by log(SSFR) $= - 3.509 - 0.7184\ x$,  where $x$=12+log(O/H).

Fig. \ref{CartoonModel} also suggests that there is a relation between the gas mass fraction and the SSFR. This indicates, regardless of the \Ms, that the intrinsic rate at which a galaxy is forming stars depends strongly on the gas mass fraction. This relationship is shown in Fig. \ref{HIgas}a. Galaxies with a high gas mass fraction show a very steep relation with SSFR. On the other hand, galaxies with a low gas mass fraction have uniformly flat and low SSFRs in which the lack of gas would prevent galaxies from forming stars. Indeed, there is an inflection point between this abrupt change at log(SSFR) $\sim-$9.9, in agreement with the dividing line seen in Fig. \ref{DependenciaBin1p1}, that motivated the cartoon model of Fig. \ref{CartoonModel}.

The relations between the gas mass fraction with the stellar mass and metallicity are shown in Figs. \ref{HIgas}b and \ref{HIgas}c. Similarly to Fig. \ref{HIgas}a, these relations show steep and flat tendencies for high and low gas mass fractions, respectively. Galaxies with high gas mass fractions have low metallicities and low--masses. This suggests, as downsizing indicates, that low--mass galaxies assemble their stars on longer timescales compared to massive galaxies. Galaxies with very low gas mass fractions, have assembled their stellar mass, exhausted their gas, and reached high stellar masses and high metallicities. As a result, galaxies with low gas mass fraction will show a flat relation against stellar mass and metallicity, as seen in Figs. \ref{HIgas}b and \ref{HIgas}c, respectively. To further refine this picture, HI measurements for massive galaxies are necessary.

\begin{figure}
\includegraphics[scale=0.28]{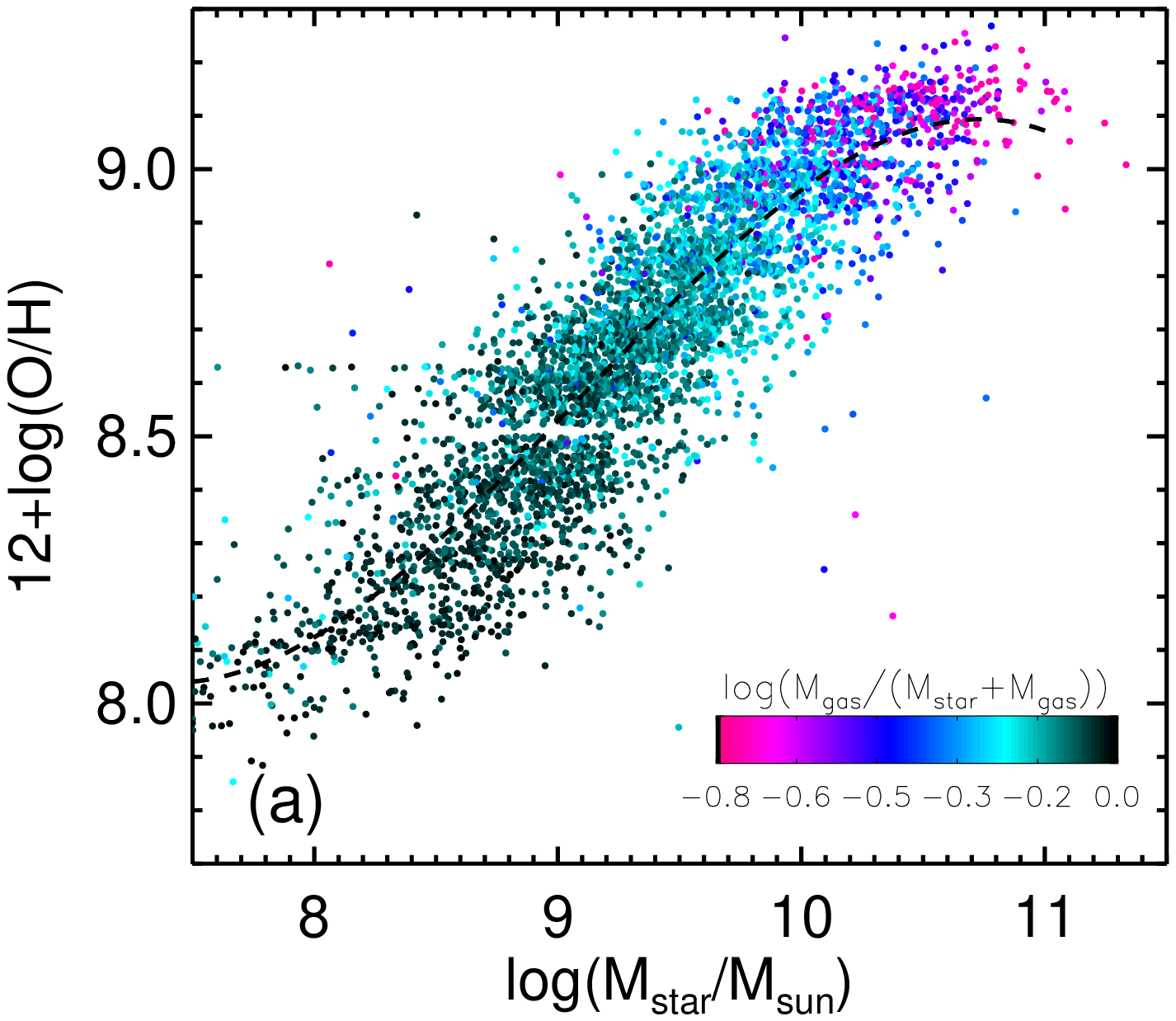}
\includegraphics[scale=0.28]{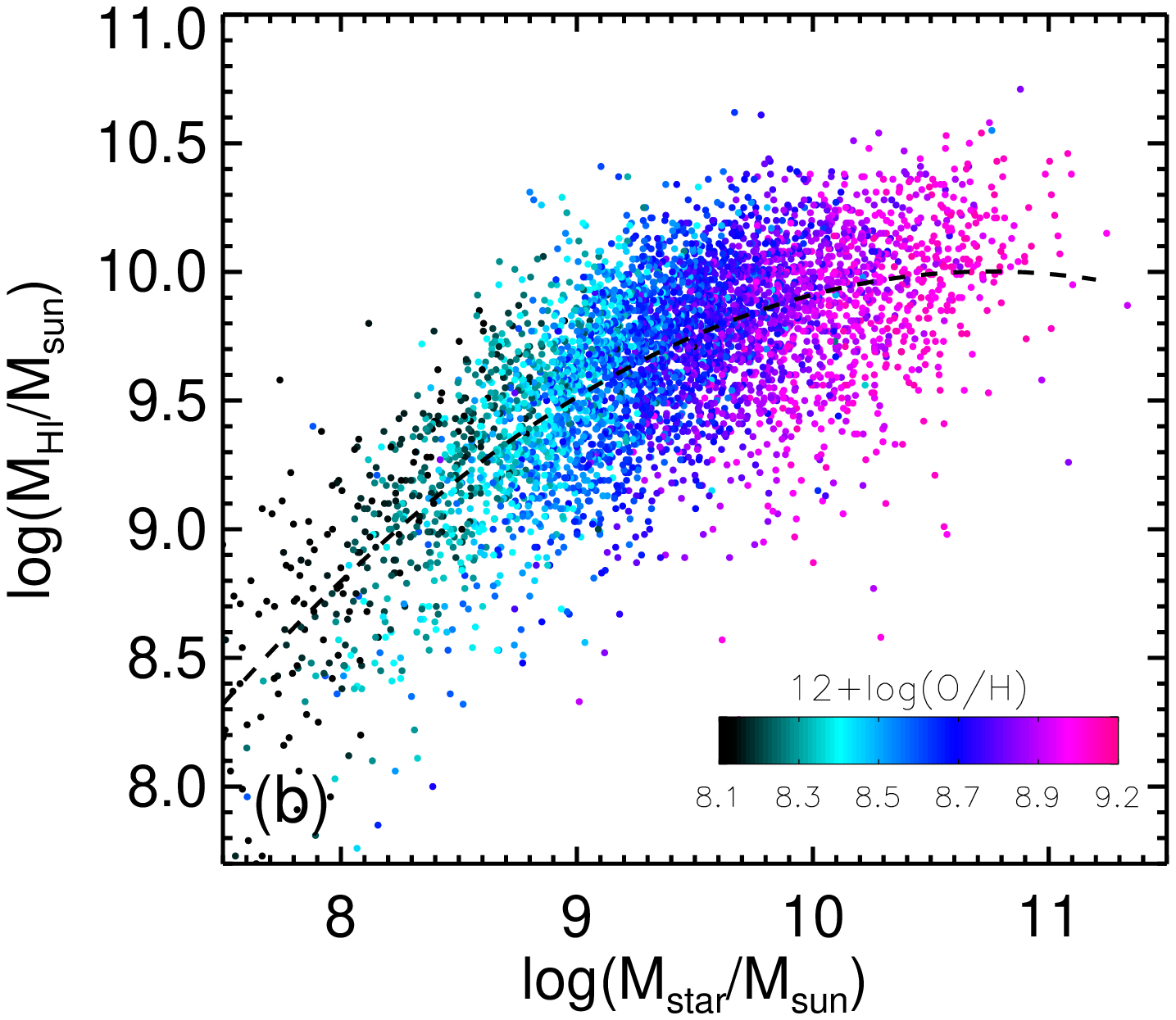}
\caption{(a) \Ms$-Z$ relation colour coded from low (magenta) to high (green)  HI/\Ms\ content. (b) log(${\rm M_{\star}/M_{\odot}}$) vs. log(${\rm M_{HI}/M_{\odot}}$) relation colour coded from low (green) to high (magenta) metallicity.}
\label{MZgas}
\end{figure}

Another way to see the influence of the gas fraction on metals is through the \Ms$-Z$ relation shown in Fig. \ref{MZgas}a. Since low--mass galaxies are actively forming stars they show a high HI gas fraction but low metallicity, while massive galaxies have finished  forming stars, exhausting their HI, and reaching high metallicities. A fit to this relation gives: 12+log(O/H) $=47.751-13.903\ x +1.5839 \ x^2-0.05815 \ x^3$, with $x$=log(${\rm M_{\star}/M_{\odot}}$).

Two important extensive variables analyzed here are the stellar mass and the total HI  in galaxies. The relationship between both is shown in Fig. \ref{MZgas}b. It is interesting to note that low--mass galaxies have a larger amount of mass in HI  than in stars, which supports the scenario described in this section. Galaxies with a higher mass in stars were more efficient at forming stars in the past, resulting in a current amount of HI that is relatively smaller, although still larger in an absolute sense, than in low-stellar-mass systems. A fit to this relation gives: log(M$_{\rm HI}$/M$_{\sun}$) $= -8.368+3.4166\ x-0.1588\ x^2$, with $x$=log(${\rm M_{\star}/M_{\odot}}$).

\section{Discussion and Conclusions}\label{Conclusion}

A key question to answer to have a general picture of this model is, why do low metallicity galaxies have either high SSFRs or low SSFRs but almost nothing in between?. In the picture presented here, galaxies evolve in a downsizing fashion with two independent mechanisms that lead to having a galaxy with low metallicities:
($i$) A high--mass galaxy will be at low metallicity today  if it exhausted its available fuel for star formation and can no longer continue to enrich its ISM.
($ii$) A  low--mass galaxy will be at low metallicity while it slowly forms stars, taking longer to enrich its ISM.
% (or loses any enriching metals into the IGM through winds perhaps, or both). SI GALAXIAS POCO MASIVAS PERDIERAN METALES POR VIENTOS, ENTONCES TAMBIEN TENDRIAN BAJA CANTIDAD DE GAS HI, COSA QUE NO ES ASI

These different scenarios lead naturally to the dearth of mid-range SSFRs for the very low metallicity systems. Staying at low-Z for massive galaxies means less and less star formation the lower the metallicity needing to be maintained. Vice-versa, for the low-mass galaxies, the metallicity stays low if the system is young and has a high amount of HI  and SSFR.
% if there are strong winds (high SSFR) or

As observed in Fig. \ref {DependenciaBin1p1} and \ref{CartoonModel}, 
% the differences in SSFR between the low and high mass branch decreases from high to low metallicity. 
within the range of stellar masses present at a given metallicity, the lower--mass end will still have higher SSFRs than  the higher--mass end, but the difference compared to the lowest metallicities will be reduced. This is because the higher mass galaxies  can have progressively more star formation and still stay at mid-range to higher metallicities, while  the lower-mass galaxies could have had a more bursty star formation in the past due to a higher amount of HI, which today would result in a high metallicity, low HI content and hence low SSFR.
% Interestingly, the fact that low mass galaxies 
% they are not losing as many metals to winds.

% An alternate explanation could be that more star formation produces more SF-driven winds but because these are low-mass galaxies, the winds drive the metals out of the galaxy into the IGM. This same mechanism in the high-mass galaxies could still be operating, but given their higher mass, they don't lose the metals they make from the winds. Nevertheless, if winds drive the metals out of the galaxy, they should also drive out the gas. According to Fig. \ref{HIgas}c however, low mass galaxies have a very high amount of HI gass compared to massive galaxies. They only galaxies that seem to be affected by winds are very low mass galaxies with \Ms$<$10$^{8.3}$ \Msun, who also show a very low metallicity.

% Therefore, we propose an ``into downsizing?", in which low mass galaxies with a high amount of HI gas will process their gas slower (and thus show low metallicities) than low mas galaxies with a lower amount of HI gas. On the other hand, massive galaxies with a high amount of HI gas will process their gas faster than massive galaxies with less amount of HI gas.

We can argue that more star formation increases the metallicity, which is sensible for massive galaxies, in which their SSFR increases as metallicity increases. It is noteworthy however, that this does not apply for low--mass galaxies, which follow the opposite behavior.

Therefore, we propose that galaxies will follow a downsizing scenario in which the  HI content is an important driver of their evolution. Low--mass galaxies with a large amount of HI  will process their gas on longer timescales and thus show low metallicities today. On the other hand, low--mass galaxies with a lower  HI might have evolved differently. These galaxies might have experienced a more bursty SF in the past that exhausted their gas and increased their metallicity.

Aditionally, a combination of the infall of pristine gas and environment could play an important role here. In this second scenario, infall can be more significant for low mass galaxies, diluting their metallicity, and provide pristine fuel for high SSFRs. It is possible, even perhaps likely, that a combination of such scenarios are at play. To explain the observations presented here, and further explore the cartoon model proposed, more detailed simulations targetted at this problem are necessary.

% The interplay of several mechanisms can have an influence to explain the cartoon model given in this letter, such as inflows, outflows, environment, and different star formation histories. At this stage, our results are purely observational, we give a couple of hypothesis but further simulation work is necessary to test different scenarios.

% On the other hand, massive galaxies with a high amount of HI gas will process their gas faster than massive galaxies with less amount of HI gas.

% In this way of thinking the "SF efficiency" that you have labelled on the cartoon model would be thought of as some combination of efficiency in producing metals from the star formation process  itself combined with efficiency in retaining them in the presence of winds.
% 
% 
% {\rm log(M_{HI}/M_{\star})}= 7.322 - 0.7435\  x
% \end{equation} where $x$=log(${\rm M_{\star}/M_{\odot}}$)

\section*{Acknowledgments}
\footnotesize{
We thank the referee for suggestions that have improved the clarity of our analysis. GAMA is a joint European-Australasian project based around a spectroscopic campaign using the Anglo-Australian Telescope. GAMA is funded by the STFC (UK), the ARC (Australia), the AAO, and the participating institutions. The GAMA website is http://www.gama-survey.org/.
Funding for the SDSS and SDSS-II was provided by the Alfred P. Sloan Foundation.
MALL thanks the ARC for funding through Super Science Fellowship FS110200023.
}

\footnotesize{
  \bibliographystyle{mn2e}

}
% \appendix

\end{document}